# METHOD OF FRACTAL DIVERSITY IN DATA SCIENCE PROBLEMS

© 2018 V.V. Vladimirov[1], E.V. Vladimirova[2]

[1] vital.vladimirov@mail.ru

[2] *Lomonosov Moscow State University, Moscow, Russia* vladimirova.elena@physics.msu.ru

The parameter (SNR) is obtained for distinguishing the Gaussian function, the distribution of random variables in the absence of cross correlation, from other functions, which makes it possible to describe collective states with strong cross-correlation of data. The signal-to-noise ratio (SNR) in one-dimensional space is determined and a calculation algorithm based on the fractal variety of the Cantor dust in a closed loop is given. The algorithm is invariant for linear transformations of the initial data set, has renormalization-group invariance, and determines the intensity of cross-correlation (collective effect) of the data. The description of the collective state is universal and does not depend on the nature of the correlation of data, nor is the universality of the distribution of random variables in the absence of data correlation. The method is applicable for large sets of non-Gaussian or strange data obtained in information technology. In confirming the hypothesis of Koshland, the application of the method to the intensity data of digital X-ray diffraction spectra with the calculation of the collective effect makes it possible to identify a conformer exhibiting biological activity.

The key steps in the derivation of the formula for the signal-to-noise ratio, which allows a quantitative comparison, are given in the article. Fractal Cantor dust or a geometric progression with an arbitrary value $0 < q < 1$ (in the classical fractal of the Cantor set $q = 2/3$) has the form:

$$F = 1 - (1-q) - (1-q)q^2 - (1-q)q^3 - (1-q)q^4 - \cdots \quad [1]$$

A method for constructing a fractal manifold is proposed. The fractal manifold for $n = 5$ of an arbitrary set of five ordered numbers $a_i$ has the form:

$$\widetilde{a_0^R}(a,5) = a_0 - (1-q)a_1 - (1-q)qa_2 - (1-q)q^2 a_3 - (1-q)q^3 a_4 - (1-q)q^4 a_0$$
$$- (1-q)q^5 a_1 - (1-q)q^6 a_2 - \cdots$$

$$\widetilde{a_1^R}(a,5) = a_1 - (1-q)a_2 - (1-q)qa_3 - (1-q)q^2 a_4 - (1-q)q^3 a_0 - (1-q)q^4 a_1$$
$$- (1-q)q^5 a_2 - (1-q)q^6 a_3 - \cdots \quad [2]$$

With each fractal cycle $m$, where $m \to \infty$, a new value $a_i$ appears from the sample of non-Gaussian data $n$ and then along the closed contour. The left and right directions of the contour are different.

In general form:

$$\widetilde{a_i^R}(a,n) = a_i - \frac{1-q}{1-q^{n+1}}\left[\sum_{k=1}^{n}\left(q^k a_{mod^1(k+1+i,n+1)}\right)\right] \quad [3]$$

Similarly, for $\widetilde{a_i^L}(a,n)$, is obtained:

$$\widetilde{a_i^L}(a,n) = a_i - \frac{1-q}{1-q^{n+1}}\left[\sum_{k=1}^{n}\left(q^{n-k} a_{mod(k+i,n+1)}\right)\right] \quad [4]$$

The sets $\{\widetilde{a_i^R}(a,n) - \widetilde{a_i^L}(a,n)\}$ and $\{\widetilde{a_i^R}(a,n) + \widetilde{a_i^L}(a,n)\}$ form fractal varieties. The expression for the signal-to-noise ratio is defined:

$$SNR(a,n) = \frac{S(a,n)}{N(a,n)} = \frac{\sum_{k=1}^{n}\left(\widetilde{a_i^R}(a,n) - \widetilde{a_i^L}(a,n)\right)^2}{\sum_{k=1}^{n}\left(\widetilde{a_i^R}(a,n) + \widetilde{a_i^L}(a,n)\right)^2} \quad [5]$$

The uniqueness of the Gaussian and Bessel functions is that the SNR [5] signal-to-noise ratio does not depend on the value of $n$. When the data are approximated by Bessel functions, the collective effect is not manifested.

Modeling non-Gaussian data with half-wave $a_i = \sin\left(\pi \frac{i}{n}\right)$, used in calculations with preliminary approximation of the data by a finite Fourier series, for sufficiently large values of $n$, the expression for the signal-to-noise ratio is:

$$S(n,q) \approx \frac{(1-q)^4(1+q)^2}{n-3} 2\pi^2(1+4q+\cdots) \quad [6]$$

---
[1] This and other formulas in the form for Mathcad.

$$N(n, q) \approx \frac{(1-q)^2(1+q)^2}{(n-3)^2} 2\pi^2(1 + 4q + \cdots) \quad [7]$$

and

$$SNR(n, q) = \big(1 - q(n)\big)^2 (n - 3) \quad [8]$$

We require the fulfillment of the invariance condition $SNR(n, q)$, which approximates strange data to Gaussian:

$$\frac{d}{dn} SNR(n, q(n)) = 0 \quad [9]$$

The solution of the differential equation has the form:

$$q(n) = 1 - \sqrt{\frac{\mu}{n - 3}} \quad [10]$$

The choice of a constant $\mu$ determines the scale of the signal-to-noise ratio.

Preliminary calculations are performed for $q = 0$ by formulas [15] − [17]. At the preliminary stage of calculations, when comparing different sets of ordered data, the critical sizes of the descriptors $n_{kr1}$, $n_{kr2}$ are obtained, which provide the maximum collective states in the data sets. Then the value $\mu = [min(n_{kr1}, n_{kr2}) - 3]$ in formula [10] and the value is calculated more accurately $SNR(max(n_{kr1}, n_{kr2}))$ taking into account the invariance ([11] − [14]) of $q$. Comparison of SNR values of different data sets is correct in the calculation performed on a single scale $\mu$. The peaks of $SNR(x_i, n)$ characterize the presence of a structure in the data variable $x$,, denote a neighborhood of the collective state. The concept of a critical or collective state is characteristic of the strange kinetics approach[1], denoting a cluster of degrees of freedom with strong correlation.

The approximation parameters of a finite Fourier series and the size of the descriptor $n$ are determined from the conditions of the maximum of the function - collection of the collective state in the system - when the ordered data are analyzed in single step.

In the matrix form, the renormalization-invariant formulas for the signal-to-noise ratio have the form:

$$SNR(a, n) = \frac{(a * Sa)}{(a * Na)} \quad [11]$$

$$S = -(matrix(n + 1, n + 1, f) - matrix(n + 1, n + 1, f)^T)^2 \quad [12]$$

$$N = [2 identity(n + 1) - (matrix(n + 1, n + 1, f) + matrix(n + 1, n + 1, f)^T)]^2 \quad [13]$$

where

$$f(i, j) = \frac{1 - q}{1 - q^{n+1}} q^{mod(j - i + n, n + 1)} \quad [14]$$

Formulas [11] − [14] are equivalent to the formulas [3] − [5] and allow programming.

In calculations from $K = n/2 + 1$ unique ordered spectrum data, a symmetric closed-loop vector is constructed:

$$a = (a_0, a_1, a_2, \cdots, a_{K-1}, a_K, a_{K-1}, \cdots, a_2, a_1) \quad [15]$$

For $q = 0$, taking into account the symmetry of the matrices $S$ and $N$, the formulas for the signal-to-noise ratio $[12] - [13]$ acquire an acceptable form for processing big data:

$$S/2 = a_0(a_0 - a_2) + a_1(a_1 - a_3) + \sum_{i=2}^{K-2} a_i(-a_{i-2} + 2a_i - a_{i-2}) + a_{K-1}(-a_{K-3} + a_{K-1})$$
$$+ a_K(-a_{K-2} + a_K) \quad [16]$$

$$N/2 = a_0(3a_0 - 4a_1 + a_2) + a_1(-4a_0 + 7a_1 - 4a_2 + a_3)$$
$$+ \sum_{i=2}^{K-2} a_i(a_{i-2} - 4a_{i-1} + 6a_i - 4a_{i+1} + a_{i+2})$$
$$+ a_{K-1}(a_{K-3} - 4a_{K-2} + 7a_{K-1} - 4a_K) + a_K(a_{K-2} - 4a_{K-1} + 3a_K) \quad [17]$$

Comparing the SNR values with the ordering scale, the scale is shifted to the left by the size of the descriptor $K$. An ordered data set, with a preliminary approximation by the finite Fourier series $k$, is analyzed by a descriptor, of size $K$, with a single step. Calculates $\sum SNR(K, k)$ by passing all the points in the data set. The objective function is defined as $max(\sum SNR(K, k))$ in the search for parameters $K$ and $k$.

As already noted, a correct comparison of the structural characteristics $SNR$ of different data sets should be carried out on a single scale $\mu$ with allowance for $q$ invariance ($[10] - [14]$). Similar to comparing measurements made in centimeters and inches.

The method is used for large sets of data obtained in good resolution, which makes it possible to increase the scale of the comparison $\mu$ with preservation of invariance. In order of magnitude, in the problem with conformers, the total number of data in the X-ray analysis spectrum is 2250 values, the optimal descriptor size for a given resolution is $K = 500$, the maximum harmonic of the finite Fourier series is $k = 3$.

The collective state in chemistry is called the flexibility or mobility of molecular fragments. Koshland's hypothesis of induced compliance with the appearance of biological activity, based on the assumption of the flexibility of the active center of the enzyme, satisfactorily explains the action of enzymes. As the substrate approaches the active center of the enzyme, a conformational restructuring occurs synchronously in the enzyme molecule, affecting a large number of degrees of freedom. The application of the computational method to the spectrum of three conformers shows a significant increase in the collective effect of the conformer, which is distinguished by its biological activity. A similar example of the collective effect is manifested in the thermomechanical curve method for polymers with different molecular weights in the region of high elasticity. The application of the method in solving the problems of data science consists in the preliminary transformation of the original non-Gaussian data and comparing the degree of cross-correlation between the data.